\documentclass[11pt]{article}           
\usepackage{booktabs}
\usepackage{amsfonts}
\usepackage{amsmath}
\usepackage{bm}
\usepackage{graphicx}
\usepackage{cite}
\usepackage{doi}
\usepackage{comment}

\title{\textbf{Simulation of multi-species flow and heat transfer using physics-informed neural networks}}

\author{Ryno Laubscher\\
Department of Mechanical Engineering \\
University of Cape Town \\
Cape Town \\
South-Africa \\
}

\begin{document}
\maketitle

The following article has been submitted to \textit{Physics of Fluids} - POF21-AR-02440.

\begin{abstract}
In the present work, single- and segregated-network PINN architectures are applied to predict momentum, species and temperature distributions of a dry air humidification problem in a simple 2D rectangular domain. The created PINN models account for variable fluid properties, species- and heat-diffusion and convection. Both the mentioned PINN architectures were trained using different hyperparameter settings, such as network width and depth to find the best-performing configuration. It is shown that the segregated-network PINN approach results in on-average 62\% lower losses when compared to the single-network PINN architecture for the given problem. Furthermore, the single-network variant struggled to ensure species mass conservation in different areas of the computational domain, whereas, the segregated approach successfully maintained species conservation. The PINN predicted velocity, temperature and species profiles for a given set of boundary conditions were compared to results generated using OpenFOAM software. Both the single- and segregated-network PINN models produced accurate results for temperature and velocity profiles, with average percentage difference relative to the CFD results of approximately 7.5\% for velocity and 8\% for temperature. The mean error percentages for the species mass fractions are 9\% for the single-network model and 1.5\% for the segregated-network approach. To showcase the applicability of PINNs for surrogate modelling of multi-species problems, a parameterised version of the segregated-network PINN is trained which could produce results for different water vapour inlet velocities. The normalised mean absolute percentage errors, relative to the OpenFOAM results, across three predicted cases for velocity and temperature are approximately 7.5\% and 2.4\% for water vapour mass fraction.
\end{abstract}
\section{Introduction}
In recent years, the application of artificial intelligence (AI) methods to thermal-fluid problems have seen growing interest from industry and research institutions alike. Some examples are: the forecasting of wind turbine power production \cite{Men2016}, the condition monitoring of power generation and refrigeration systems \cite{Hundi2020} \cite{Haffejee2021} \cite{Zhang2020}, surrogate model development for design optimisation of solid body temperature distribution \cite{Yang2019}, co-simulation of surrogate and computational fluid dynamic (CFD) models \cite{Singh2018}, prediction of jet diffusion flame species and temperature fields \cite{Laubscher2021} and combustion control system optimisation using reinforcement learning \cite{Cheng2018}. A common application, as seen from the mentioned examples, is the use of neural networks or other machine learning models to develop computationally efficient surrogate models. Surrogate models can be used for design optimisation, multi-scale simulation, inverse modelling to find unknown boundary conditions and fast-simulation for performance monitoring \cite{Hennigh2020}. Creating surrogate models of thermal-fluid processes using experimental and/or simulation data has drawbacks. The former suffering from the sparsity of measurement data and the latter of from the high cost associated with generating simulation data \cite{Raissi2019}, especially, using traditional CFD techniques such as the finite volume method (FVM). Physics-informed neural networks (PINN), developed by Raissi et al. \cite{Raissi2019}, aims at overcoming these challenges by incorporating known knowledge such as governing physics equations into the learning structure of the neural network models. This enables PINNs to resolve the governing equations without the need for simulation results or experimental data as inputs. 

In the present work, the accuracy of PINNs to resolve the forward problem for steady-state multi-species convective and diffusive flow and heat transfer predictions will be evaluated using a simple case study. The results of the PINN models will be compared to simulation data taken from a traditional CFD simulation model prepared in OpenFOAM, an open-source fluid flow simulation library. 

Recently many authors have applied PINNs to investigate various thermal-fluid problems ranging from testing the accuracy of PINNs compared to traditional simulation approaches to increasing numerical solving efficacy of PINNs \cite{Willard2020}. For example, Rao et al. \cite{Rao2020} proposed a mixed-variable PINN approach that eliminates the need to calculate higher-order derivatives in the loss function of the network and showcased the performance improvements using a simple 2D case study of fluid flow over a cylinder. Cai et al. \cite{Cai2021} applied PINNs to various heat transfer problems solving both the momentum and energy transport governing equations. The trained PINNs showed exceptional ability to tackle the inverse problem by inferring unknown thermal boundary conditions for mixed convective flow. The authors also applied PINNs for heat sink design optimisation, to showcase the industrial application of the technique. Mao et al. \cite{Mao2020} investigated the accuracy of PINNs to resolve forward and inverse problems in high-speed flows governed by the Euler equation. Jagtap et al.\cite{Jagtap2020} proposed an adaptive activation function that aims at accelerating the convergence speed of PINN training. The authors showed that applying the adaptive function in modelling the Klein-Gordan, non-linear Burger and Helmholtz equations yielded faster convergence compared to the traditional constant activation functions used in deep learning. PINNs are notoriously expensive to train for long-time periods and is more suited for short-time integration, therefore, Meng et al.\cite{Meng2020} proposed a para-real PINN approach which decomposes the problem into many short-time integration networks. The authors applied the proposed techniques to a 1D Burgers equation problem and a 2D reaction-diffusion problem, showing significant improvement in convergence time during PINN training. Mishra and Molinaro \cite{Mishra2021} used PINNs to accurately resolve the radiation transport equation. The authors applied PINNs in both the forward and inverse mode to various radiation transport problems.

In the present work, the accuracy of PINNs to resolve the steady-state momentum, species and energy equations simultaneously will be explored. Two PINN modelling approaches will be implemented. The first, which is the typical approach, uses a single network to resolve all the physics and will be called the PINN-1 approach. The second approach, called PINN-3, uses three networks to solve the governing equations and transports the relevant solution variables between the PINNs during training. As a case study a simple 2D dry air humidification duct flow problem will be solved and the PINN results compared to the CFD simulation results. Furthermore, a hyperparameter search is performed to compare the achieved loss values between the PINN-1 and PINN-3 approaches for different depth and width neural networks. In the present work, PyTorch \cite{Mazza2021} library is utilised to create and train the various PINNs. According to the knowledge of the authors, no previous work has investigated the application of PINNs to simulate mass, momentum, species and energy transport. Furthermore, the application of a segregated-PINN approach, one for each set of physics equations, also has not been studied before.

\section{Material and methods}
\subsection{Overview of physics-informed neural networks}

In a 2D steady-state PINN framework, multilayer perceptron (MLP) \cite{Goodfellow2017a} neural networks are used to predict the solution values $\bm{\phi}^*(\bm{x})$ of selected partial differential equations ($*$ denotes network predicted values) \cite{Raissi2019}. These variables can represent various physical transported fields such as $i^{th}$ species mass fraction $Y_i$, X-velocity $u$ or temperature $T$. The inputs to these MLP networks are spatial coordinates $\bm{x}(x,y)$ located within the computational domain $\mathcal{D}$. 

The steady-state partial differential equations (PDEs) approximated by the MLP typically have the following generic structure:

\begin{equation} \label{eq:1}
	\prod \{ \bm{\phi} (\bm{x}) \} = 0, \hspace{3mm} \bm{x} \in \mathcal{D}
\end{equation}
\begin{equation} \label{eq:2}
	\bm{\phi}(\bm{x}) = \bm{b}^{bc}(\bm{x}), \hspace{3mm} \bm{x} \in \partial \mathcal{D}
\end{equation}

In \eqref{eq:1}, $\prod \{ \cdot \}$ is the linear and nonlinear differential operator which represents a wide range of PDE quantities such as advection, diffusion and source terms. In \eqref{eq:2}, $\bm{b}^{bc}(\cdot)$ is the boundary condition functions and $\partial \mathcal{D}$ denotes the boundary domains, which is simply a collection of $x$ and $y$ coordinates located on the boundary edges. To calculate the linear and nonlinear quantities ($\prod \{ \bm{\phi}^*(\bm{x}) \}$) required to construct the desired PDEs from the PINN outputs $\bm{\phi}^*$, automatic differentiation \cite{Kochenderfer2019} through the network graph structure is utilised, which simplifies the estimation of network output gradients with respect to the input spatial dimensions. 

To ensure the network accurately approximates the desired PDEs and associated boundary conditions, a combined loss value $\mathcal{J}_L$ is minimised. The calculation of this combined loss value can be seen in \eqref{eq:3} below, where $K$ is the number of PDEs being solved and $J$ the number of boundaries located in the domain $\mathcal{D}$.

\begin{equation} \label{eq:3}
	\mathcal{J}_L = \sum_k^K J_r^k + \sum_k^K \sum_j^J J_{b,j}^k
\end{equation}

 The combined loss value for a 2D steady-state PINN consists of two parts. The first being the residuals of the modelled PDEs, shown in \eqref{eq:4} for the $k^{th}$ PDE. The second is the boundary conditions losses and is calculated for the $k^{th}$ PDE and $j^{th}$ boundary as shown in \eqref{eq:5}. In the equations below, $N_{i}$ denotes the number of sampling points within the domain and $N_{b,j}$ the number of sampling points located on boundary $j$.

\begin{equation} \label{eq:4}
	\mathcal{J}_{r}^{k} = \frac{1}{N_{i}} \sum_{i = 1}^{N_{i}} \left( \prod \{ \bm{\phi}^*_k(\bm{x}_i) \} \right)^2
\end{equation}

\begin{equation} \label{eq:5}
	\mathcal{J}_{b,j}^{k} = \frac{1}{N_{b,j}} \sum_{i=1}^{N_{b,j}} \left( \bm{\phi}^*_k(\bm{x}_i) - \bm{b}_{j}^k(\bm{x}_i) \right)^2
\end{equation}

As seen in the equations above, both the loss quantities $\mathcal{J}_r^k$  and $\mathcal{J}^k_{b,j}$ uses the predicted solution field variables $\bm{\phi}^*(\bm{x})$ as input. To calculate these PINN outputs, the spatial coordinates of each selected point in $\mathcal{D}$ and $\partial \mathcal{D}$ is propagated through the MLP network structure to generate a prediction. The propagation process, entails calculating each network layer output $\bm{h}_l$, which involves multiplying the previous layer output $\bm{h}_{l-1}$ with the layer weights $\bm{W}_l$, adding the biases $\bm{b}_l$ and passing the resultant matrix into the layer activation function $\sigma_l$, as shown in \eqref{eq:6}. There are many different types of activation functions proposed in literature, in the present work the hidden layers of the PINNs use hyperbolic-tanh activation functions and the final layer a linear activation function \cite{Geron2017a}. For the initial network layer the input is the spatial coordinates for the boundaries $\bm{x}_{b,j} = [(x_1 , y_1), (x_2, y_2),...,(x_{N_{b,j}}, y_{N_{b,j}})]$ and internal points 

$\bm{x}_i = [(x_1 , y_1), (x_2, y_2),...,(x_{N_i}, y_{N_i})]$ and the resultant final layer ($L$) output is simply the PINN output prediction for a given set of sampling points, $\bm{h}_L = \bm{\phi}^*(\bm{x})$.

\begin{equation} \label{eq:6}
	\bm{h}_l = \sigma_l \left( \bm{h}_{l-1} \cdot \bm{W}_l + \bm{b}_l \right)
\end{equation}

To ensure the predicted outputs accurately approximate the solution of the selected PDEs, the weights and biases of the MLP network layers are optimised to minimise the combined loss value $\mathcal{J}_L$. The minimisation of the loss value is achieved using gradient-based optimisation, which requires the gradients of the weights and biases ($\nabla_{\bm{W}} \mathcal{J}_L$, $\nabla_{\bm{b}} \mathcal{J}_L$) with respect to the loss to be calculated. These gradients are calculated similarly to the network output-input gradients using automatic differentiation. Once the gradients are known, the trainable parameters can be iteratively updated using a first-order optimisation algorithm. In the present work, the Adam algorithm \cite{Kingma2015} is used to update the weights and biases of the MLP network, using a training schedule which decreases the learning rate parameter $\eta$ after a set number of training iterations, more on this in section \ref{sec:train}.

\subsection{Case study problem}

In the current work, the propagation of water vapour into dry air flowing in a duct is simulated using the PINN framework with the PyTorch library and traditional FVM using the OpenFOAM \cite{Weller1998} software suite. The domain under consideration is a simple 2D rectangular cartesian domain of dimensions $0.5 \times 0.1 \; [m]$ with a dry air inlet (inlet), mixture outlet (outlet), water vapour inlet (bot wall) and constant temperature solid wall (top wall) boundary conditions as seen in figure 1.

\begin{figure}[h] \label{fig:1}
	\centering
	\includegraphics[scale=0.325]{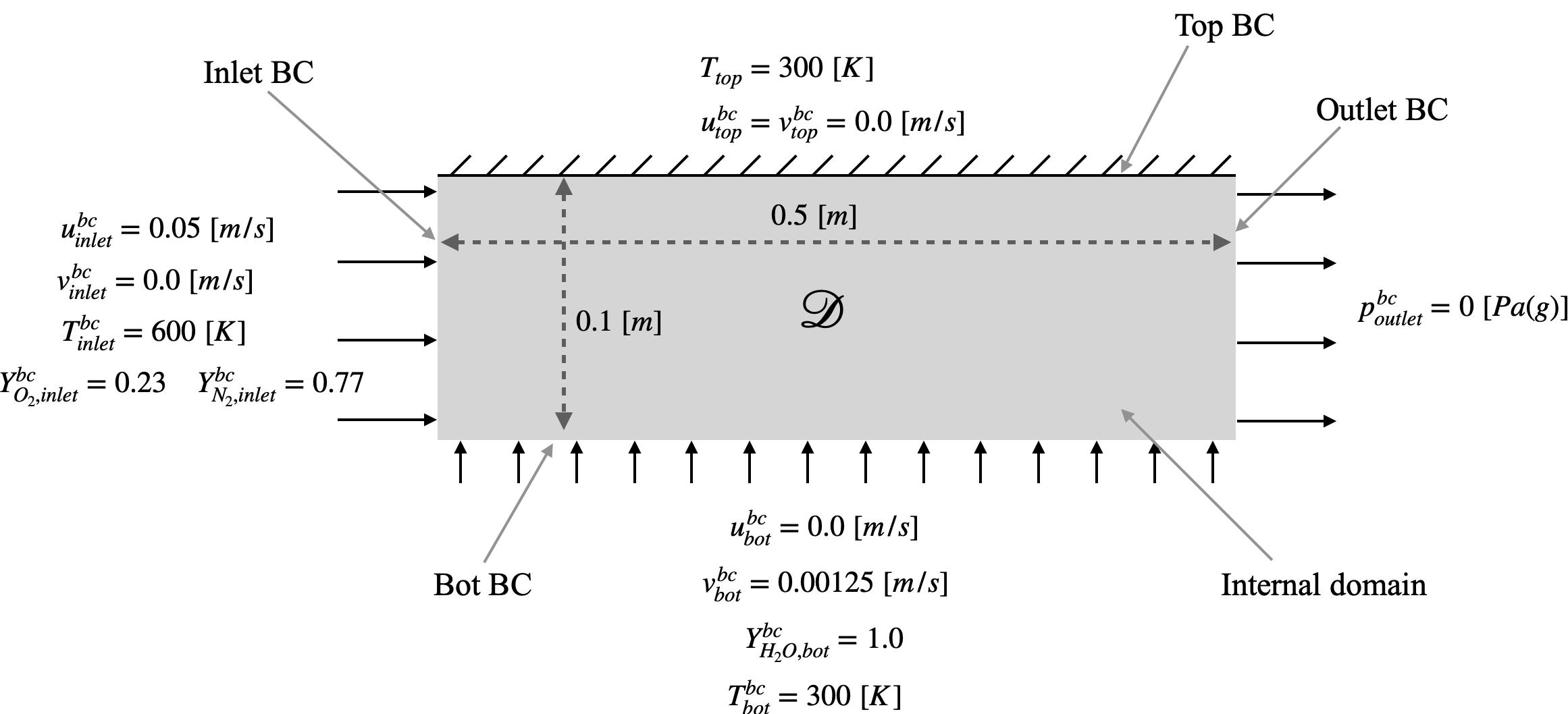}
	\caption{Computational domain showing boundary conditions}
\end{figure}

To solve the laminar steady-state transport of mass, momentum, species and energy throughout the domain requires the simultaneous solution of the following PDEs \cite{ANSYSInc.2015}:

Mass transport:
\begin{equation} \label{eq:7}
	\nabla \cdot \bm{u} = 0
\end{equation}
Momentum transport:
\begin{equation} \label{eq:8}
	\rho \nabla \cdot \left( \bm{u} \bm{u} \right) = \mu \nabla \cdot \left( \nabla \bm{u} \right) - \nabla p \quad \text{where} \; \bm{u} = \left( u,v \right)
\end{equation}
Species transport:
\begin{equation} \label{eq:9}
	\rho \nabla \cdot \left( \bm{u} Y_i \right) = \rho D_m \nabla \cdot \left( \nabla Y_i \right) \quad \text{where} \; i=O_2, N_2, H_2O
\end{equation}
Energy transport:
\begin{equation} \label{eq:10}
	\rho \nabla \cdot \left( \bm{u} c_{P,m}T \right) = \nabla \cdot \left( \lambda \nabla T \right) + \rho D_m \nabla \cdot \left( \sum_{i=1}^3 h_i \nabla Y_i \right)
\end{equation}

For the PINN and FVM models, the density field and the relevant transport equations are solved assuming the fluid mixture is an incompressible ideal gas  \cite{Batchelors2000}, therefore, the density is a function of the local gas temperature and mixture composition, only. This is a valid simplification seeing as $\beta \theta \frac{\kappa}{L u} \approx 6 \times 10^{-6}$ ($\beta$ thermal expansion coefficient, $\theta$ non-dimensional temperature, $\kappa$ thermal diffusivity, $L$ characteristic length of the domain and $u$ freestream velocity), which means the heat conduction is relatively small and does not significantly affect the density field. The specific heat capacity of the gas mixture $c_{P,m}$ is calculated using a mass-weighted mixture-averaged approach where the individual species heat capacities $c_{P,i}$ are calculated using real gas polynomials extracted from the CoolProp fluid property database \cite{Bell2014}. The species enthalpies $h_i$ was also calculated using the mentioned property database. The thermal conductivity and viscosity are assumed to be constant values of $\lambda = 0.02 \; [\frac{W}{m \cdot K}]$ and $\mu = 1.779 \times 10^{-5} \; [Pa \cdot s]$ respectively. The constant dilute approximation is assumed for the mixture molecular diffusion coefficient, $D_m = 2.88 \times 10^{-5} \; [\frac{m^2}{s}]$.

\subsection{Model development}
\subsubsection{Normalised physics equations}
The magnitudes of the solution variables in momentum, heat and species transport simulations typically have large differences. For example, in the present work the temperature field values will be between $300 \rightarrow 600 \; [K]$ and the velocity field between $0 \rightarrow 0.1 \; [m/s]$. This large difference between the momentum and temperature scalars will result in the loss values of the energy PDE ($\mathcal{J}^T_r + \sum_j \mathcal{J}^T_{b,j}$)  contributing a larger portion to the combined loss value than the momentum PDE losses ($\mathcal{J}^u_r + \sum_j \mathcal{J}^u_{b,j}$). The result of this is the biasing of the optimiser during the training of the network parameters ($\bm{W}, \bm{b}$). The optimiser will, therefore, favour the energy residuals and boundary conditions, which leads to extended network training times and inaccurate results. To circumvents this, the transport equations are normalised before being implemented in the PINN framework. The normalised inputs and solution variables are defined as:

\begin{equation} \label{eq:11}
	x^*=\frac{x}{L}, \; y^*=\frac{y}{L}, \; u^*=\frac{u}{u_\infty}, \; v^*=\frac{v}{u_\infty}, \; p^* = \frac{p}{\rho u^2_\infty}, \; T^* = \frac{T}{T_\infty}
\end{equation}

In \eqref{eq:11}, $^*$ denotes the non-dimensional variable. The spatial coordinates $x,y$ are normalised using the characteristic length $L = 0.5 \; [m]$ of the domain. The $x$ and $y$ velocity components $u,v$ are normalised using the inlet X-velocity, $u_{inlet} = 0.05 \; [m/s]$. The static pressure field $p$ is normalised using the maximum local dynamic pressure $\rho u_\infty^2$. Using the above normalised quantities and the transport equations shown in \eqref{eq:7} $\rightarrow$ \eqref{eq:10}, the equations solved for selected internal coordinates by the PINNs are written as follows: \\

Mass:
\begin{equation} \label{eq:12}
	\frac{\partial u^*}{\partial x^*} + \frac{\partial v^*}{\partial v^*} = 0
\end{equation}

X-momentum transport:
\begin{equation} \label{eq:13}
	u^* \frac{\partial u^*}{\partial x^*} + v^* \frac{\partial u^*}{\partial y^*} = \frac{\partial \sigma_{11}^*}{\partial x^*} + \frac{\partial \sigma^*_{12}}{\partial y^*}
\end{equation}

Y-momentum transport:
\begin{equation} \label{eq:14}
	u^* \frac{\partial v^*}{\partial x^*} + v^* \frac{\partial v^*}{\partial y^*} = \frac{\partial \sigma_{12}^*}{\partial x^*} + \frac{\partial \sigma^*_{22}}{\partial y^*}
\end{equation}

with 

\begin{align*}
		\sigma_{11}^*  = -p^* + \frac{2}{Re} \frac{\partial u^*}{\partial x^*}, \qquad
		\sigma_{22}^* = -p^* + \frac{2}{Re} \frac{\partial v^*}{\partial y^*} \\ \\
		\sigma_{12}^*  = \frac{1}{Re} \left( \frac{\partial u^*}{\partial y^*} + \frac{\partial v^*}{\partial x^*} \right), \qquad
		p^* = -\frac{\sigma_{11}^* + \sigma_{22}^*}{2}
\end{align*}

$i^{th}$ Species transport:

\begin{equation} \label{eq:15}
	u^* \frac{\partial Y_i}{\partial x^*} + v^* \frac{\partial Y_i}{\partial y^*} = \frac{\partial J^*_x}{\partial x^*} + \frac{\partial J^*_y}{\partial y^*} 
\end{equation}

with

\begin{align*}
	J^*_x = \frac{1}{ReSc} \frac{\partial Y_i}{\partial x^*}, \qquad J^*_y = \frac{1}{ReSc} \frac{\partial Y_i}{\partial y^*}
\end{align*}

Energy transport:

\begin{equation}
\begin{split}
& \hspace{37mm} u^* \frac{\partial T^*}{\partial x^*} + v^* \frac{\partial T^*}{\partial y^*} = \frac{\partial q^*_x}{\partial x^*} + \frac{\partial q^*_y}{\partial y^*} +  \\ & \frac{1}{ReSc} \left( \sum_{i=1}^3 \frac{c_{P,i}}{c_{P,m}} \left( \frac{\partial Y_i}{\partial x^*} \frac{\partial T^*}{\partial x^*} + \frac{\partial Y_i}{\partial y^*} \frac{\partial T^*}{\partial y^*}\right) + \sum_{i=1}^{3} \frac{h_i}{c_{P,m}} \left( \frac{\partial}{\partial x^*}\left( \frac{\partial Y_i}{\partial x^*} \right) + \frac{\partial}{\partial y^*} \left( \frac{\partial Y_i}{\partial y^*} \right) \right) \right)
	\end{split}
\end{equation}
with 
\begin{align*}
	q^*_x = \frac{1}{RePr} \frac{\partial T^*}{\partial x^*}, \qquad q^*_y = \frac{1}{RePr} \frac{\partial T^*}{\partial y^*}	
\end{align*}

The various non-dimensional numbers used above are: $Re = \frac{\rho u_\infty L}{\mu}$, $Pr = \frac{c_{P,m} \mu}{\lambda}$ and $Sc = \frac{\mu}{\rho D}$. 

To ensure the physics equations are enforced in the computational domain by the learnt network parameters, requires the minimisation of the residual loss values, namely $\mathcal{J}_r^{m^*}$ (mass), $\mathcal{J}_r^{u^*}$ (X-momentum), $\mathcal{J}_r^{v^*}$ (Y-momentum), $\mathcal{J}_r^{Y_i}$ (species) and $\mathcal{J}_r^{T^*}$ (energy). To calculate the residual loss values entails writing the above normalised equations with all quantities on one side of the equals sign then taking the square average over all $i$ internal points in the domain. For example, the residual losses for the X-momentum and $i^{th}$ species transport equations is calculated as:

\begin{equation}
	\mathcal{J}^{u^*}_r = \frac{1}{N_i} \sum_{i=1}^{N_i}	\left( u^* \frac{\partial u^*}{\partial x^*} + v^* \frac{\partial u^*}{\partial y^*} - \frac{\partial \sigma_{11}^*}{\partial x^*} - \frac{\partial \sigma^*_{12}}{\partial y^*} \right)_i^2
\end{equation}

\begin{equation}
		\mathcal{J}^{Y_i}_r = \frac{1}{N_i} \sum_{i=1}^{N_i} \left( u^* \frac{\partial Y_i}{\partial x^*} + v^* \frac{\partial Y_i}{\partial y^*} - \frac{\partial J^*_x}{\partial x^*} - \frac{\partial J^*_y}{\partial y^*} \right)_i^2
\end{equation}

An additional residual loss value is used to ensure the summation of all the species mass fractions at every point in the domain equals to unity, $\mathcal{J}^{\sum Y_i}_r = \frac{1}{N_i}\sum_{i=1}^{N_i} \left( Y_{i,H_2O} + Y_{i,O_2} + Y_{i,N_2} - 1.0 \right)^2$. 

To eliminate the need to calculate higher order derivatives using automatic differentiation, the mixed-variable approach of Rao et al. \cite{Rao2020} was implemented, which requires the various flux variables, such as $\sigma^*_{11}$, $q^*_{x}$ and $J^*_{i,x}$, to be also calculated as network outputs and enforced using residual losses (these calculated variables are indicated with a superscript $p$). The additional residual losses are: $\mathcal{J}_r^{\sigma^*_{11}}$, $\mathcal{J}_r^{\sigma^*_{12}}$, $\mathcal{J}_r^{\sigma^*_{22}}$, $\mathcal{J}_r^{q^*_{x}}$, $\mathcal{J}_r^{q^*_{y}}$, $\mathcal{J}_r^{J^*_{i,x}}$ and $\mathcal{J}_r^{J^*_{i,y}}$. Examples of the flux residuals for $\sigma^*_{11}$ and $q^*_{x}$ are shown below.

\begin{equation}
	\mathcal{J}^{\sigma^*_{11}}_r = \frac{1}{N_i} \sum_{i=1}^{N_i} \left( \sigma^{*,p}_{11,i} - \sigma^*_{11} (\text{eq.14}) \right)
\end{equation}

\begin{equation}
	\mathcal{J}^{q^*_{x}}_r = \frac{1}{N_i} \sum_{i=1}^{N_i} \left( q^{*,p}_{x,i} - q^*_{x} (\text{eq.16}) \right)
\end{equation}

The residual loss values are only enforced at internal points and not at the boundary edges. Therefore, boundary closure is required before a solution can be found. This is achieved by, as mentioned, defining boundary loss functions which similarly to the residual losses are minimised during training of the network. Minimising the boundary loss functions will ensure the various inlet, outlet and wall boundary conditions are imposed. The estimation of the boundary loss functions entails calculating the solution variables at the boundary edges and forcing the calculated values to the desired boundary condition values, shown in figure 1. For example, the inlet boundary velocity loss functions are shown below in \eqref{eq:21}.

\begin{equation} \label{eq:21}
	\begin{split}
			\mathcal{J}_{b,inlet}^u = \frac{1}{N_{b,inlet}} \sum_{i=1}^{N_{b,inlet}} \left( u^*_{i,inlet} -\frac{u_{inlet}^{bc}}{u_\infty} \right)^2,  \\
	\mathcal{J}_{b,inlet}^v = \frac{1}{N_{b,inlet}} \sum_{i=1}^{N_{b,inlet}} \left( v^*_{i,inlet} -\frac{v_{inlet}^{bc}}{u_\infty} \right)^2
	\end{split}
\end{equation}

In the next sections, the layout and training algorithm of the PINN-1 and PINN-3 models will be discussed and how the above residual loss and boundary loss functions are implemented.

\subsubsection{Physics-informed network models}
Two implementations of PINNs to model multi-species flow and heat transfer is investigated in the current work. The first, as mentioned, uses a single network to predict the mass, momentum, species and energy equation quantities, whereas, the second approach uses a PINN for each set of physics PDEs being solved (1 for mass and momentum PDEs, 1 for all the species PDEs and 1 for the energy PDEs).

Figure 2 below, shows a schematic of the PINN-1 model with an overview of the calculated quantities. The PINN-1 layout is the archetypical PINN configuration, pioneered by \cite{Raissi2019} and is trained using a single optimiser routine to solve all the physics equations simultaneously. In the present work, the MLP network accepts the $x^*,y^*$ coordinates and outputs the 18 physical quantities. These 18 quantities evaluated at the various boundaries and internal points are then used to calculate the combined loss value. During the training of the network parameters, the variable fluid properties ($\rho$ and $c_{P,m}$) are updated for each internal points using the predicted temperature and species composition for each training iteration. The updated fluid properties are used along with the predicted physical quantities to estimate the loss values. 

\begin{figure}[h] \label{fig:2}
 \hspace*{-0.75cm}
	\includegraphics[scale=0.25]{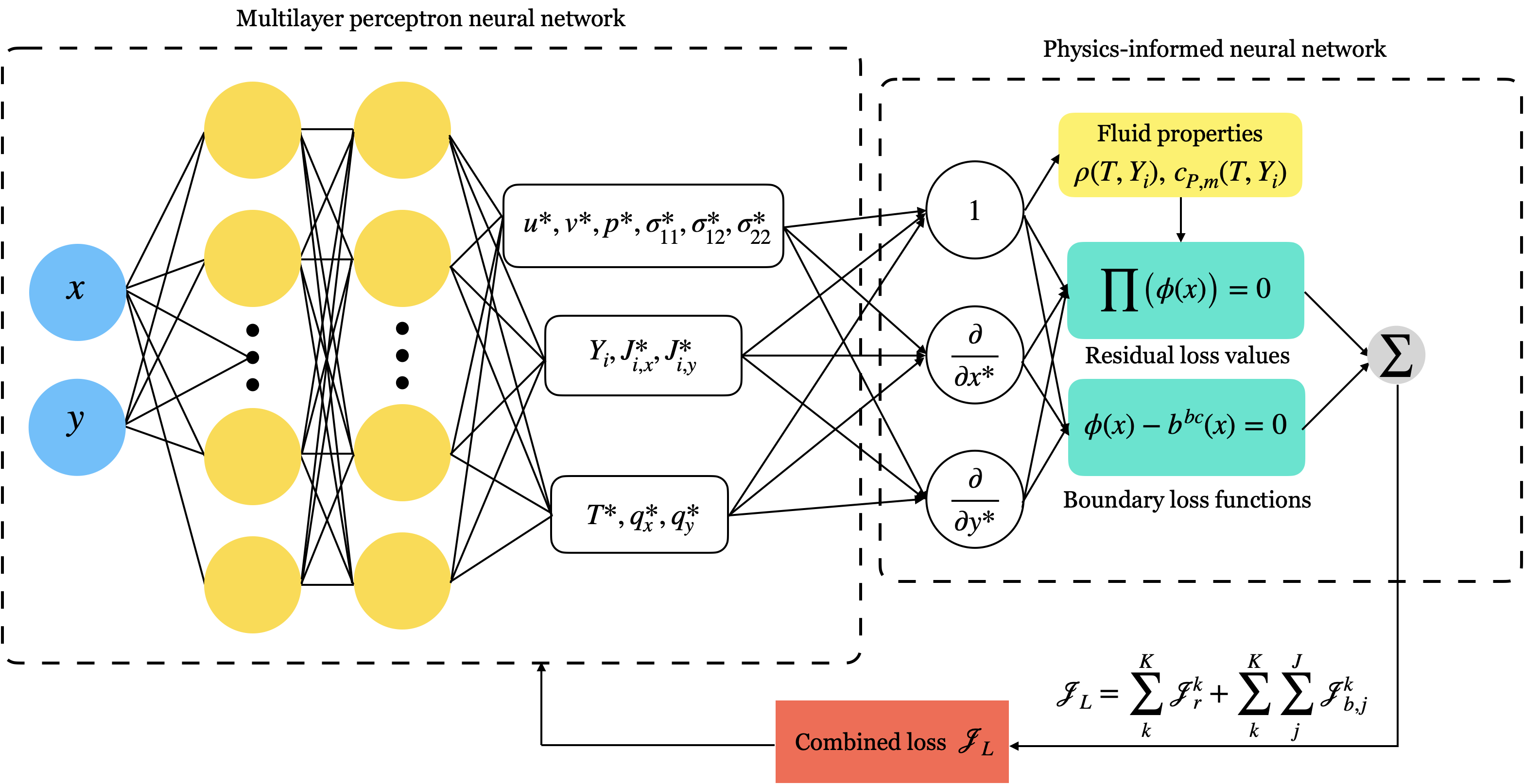}
	\caption{Physics-informed neural network layout for current problem using a single MLP network (PINN-1)}
\end{figure}

Figure 3, shows a schematic of the PINN-3 configuration. From the schematic, it is seen that the normalised spatial coordinates are fed to the three MLP networks which then predicts the various physical quantities. The concept of using three different solvers, one for each set of physics equations, is inspired by the traditional approach of multiple solvers for the sets of linear equations in FVM \cite{Moukalled2016}. For example, in FVM the pressure field is solved with GAMG, velocity field with a preconditioned Bi-conjugate gradient solver and the temperature and species fields using a Gauss-Seidel smooth solver. 

\begin{figure}[h!] \label{fig:3}
 \hspace*{-1.75cm}
	\includegraphics[scale=0.25]{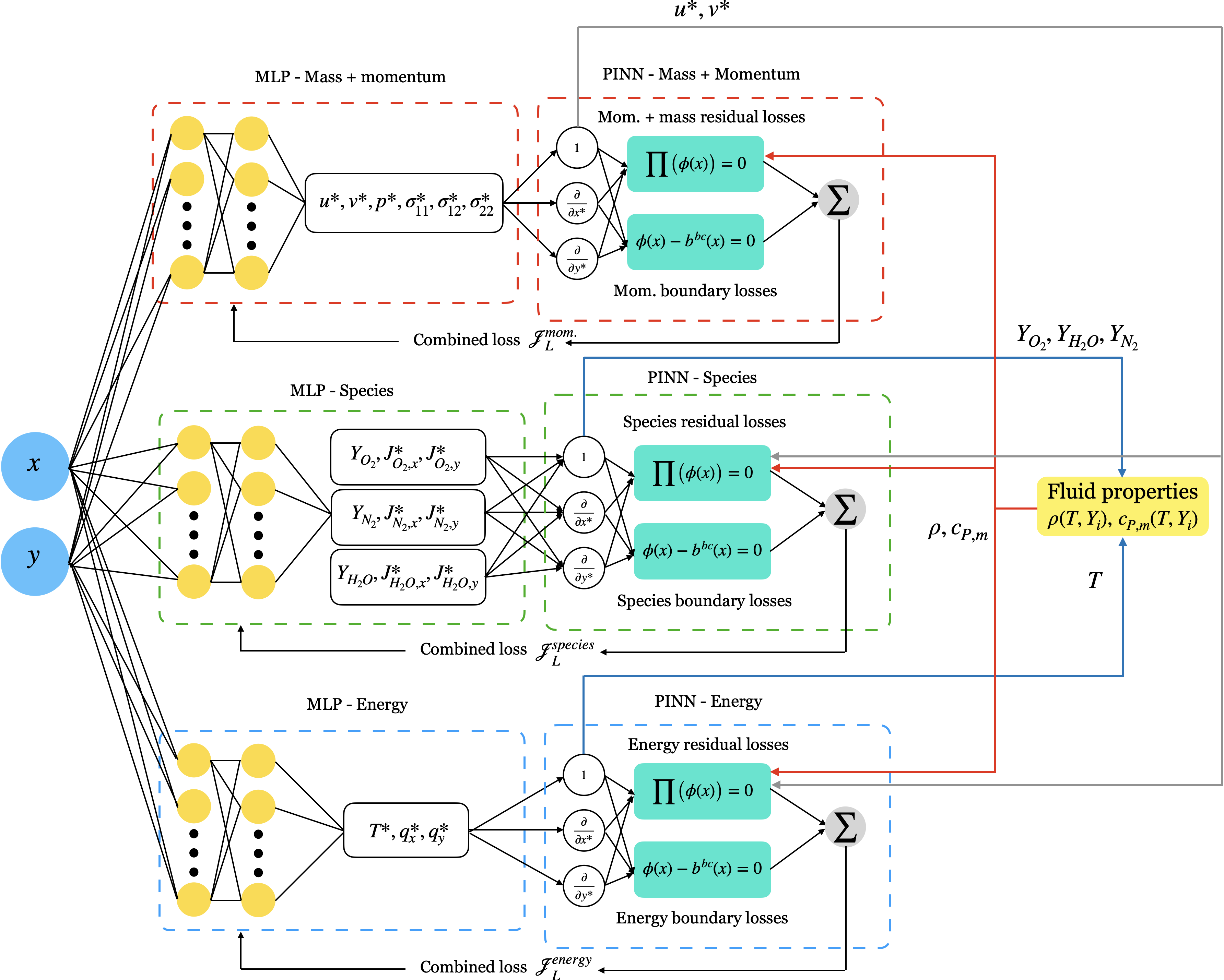}
	\caption{Physics-informed neural network layout for current problem using a three MLP networks (PINN-3)}
\end{figure}

 For the PINN-3 approach, the predicted values of each MLP network is used to calculate the residual and boundary condition losses of the PINNs, similarly to the PINN-1 model. The losses are then used to update the weights and biases of each MLP network. Automatic differentiation is performed for each network to calculate the gradients of the network-specific combined loss with respect to that network trainable parameters. The three sets of loss gradients are then used with three Adam optimisers to update the weights and biases of each network individually.

As seen in figure 3, some of the predicted variables are directly or indirectly shared between the PINNs. The transport of these predicted quantities between the PINNs is required to calculate the residual losses of each PINN. For example, the predicted normalised velocities are fed to the residual loss calculation blocks of the species and energy transport PINNs to calculate the convective quantities of the PDEs. Similarly, the temperature and composition at each internal point in the domain are sent to the fluid properties function which then returns the density field to each network. The specific heat capacity is additionally returned by the fluid properties function to the energy PINN. The training strategy of the PINN-3 approach incorporates the inter-network sharing of predicted physical quantities. Figure 4 shows, a flow diagram of the training strategy. The training strategy starts by specifying the learning rates of the three Adam optimisers, and initialisation of the variable fluid properties for each internal points in the domain. For the first 100 training iterations, the fluid properties are kept constant using dry air composition at $300 \; [K]$ as inputs. After the first 100 iterations, the fluid properties are updated every training cycle using the various internal point temperatures and species compositions. Once the fluid properties are calculated and passed to the respective PINN instances, the mass and momentum PINN combined loss and the MLP parameter-loss gradients are calculated. The parameter gradients are applied by the Adam optimiser of the mass and momentum PINN to update the weights and biases. Once updated the mass and momentum PINN is used to calculate the normalised velocity field and the calculated values are shared with the species- and energy-PINNs. A similar process is followed for the species- and energy-PINNs, which uses separate Adam optimisers to update their training parameters. The estimated species composition and temperature fields are stored and used at the beginning of the next training epoch as inputs to the fluid property calculation.

In the next section, the training results of the PINN-1 and PINN-3 models will be discussed along with the sensitivity of the combined loss values on MLP network hyperparameters such as network width and depth.

\begin{figure}[h!] \label{fig:4}
	\includegraphics[scale=0.25]{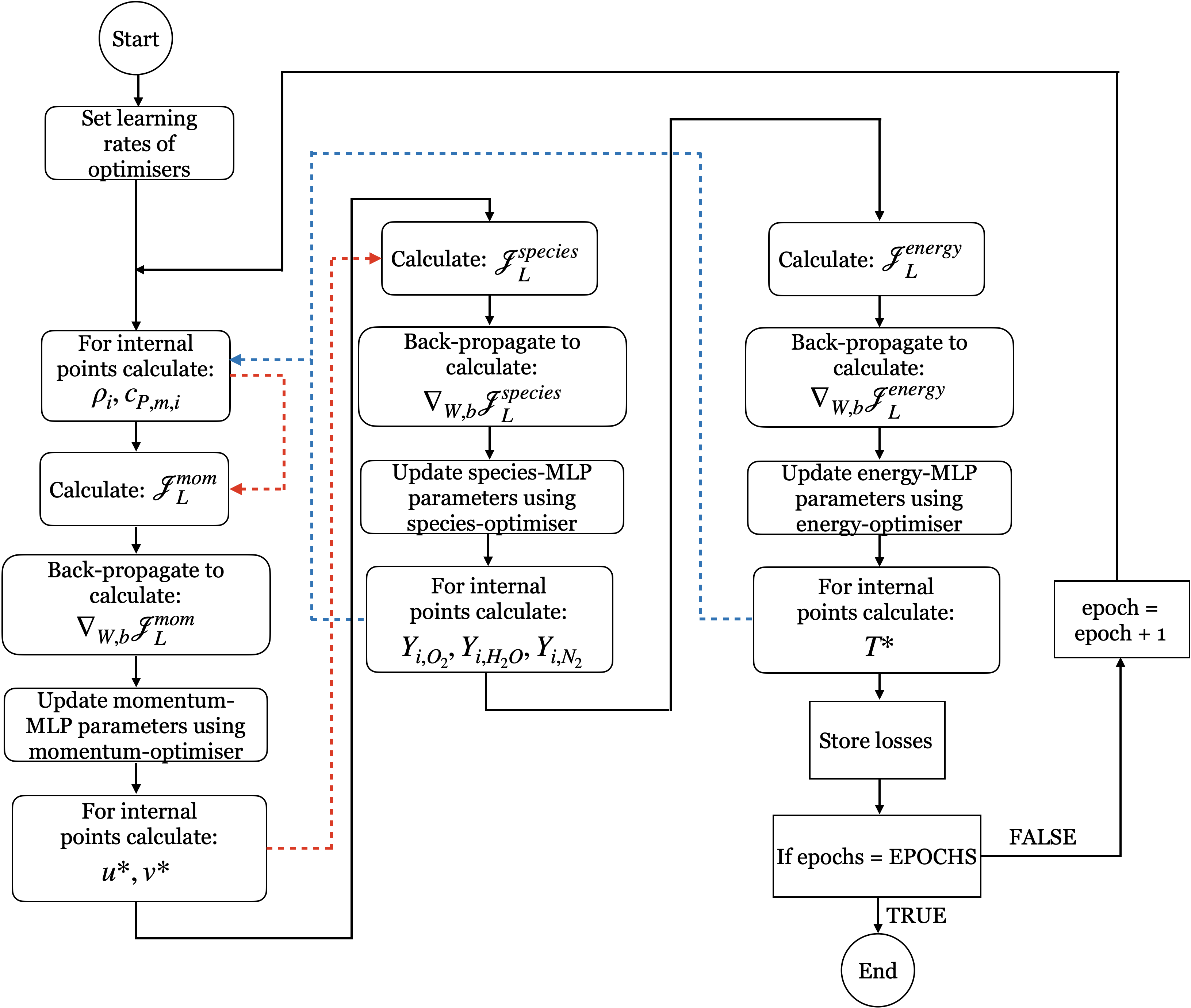}
	\caption{Training strategy of PINN-3 model}
\end{figure}

\section{Results and discussion}
\subsection{Training and hyperparameter search results} \label{sec:train}
To gauge the effect of MLP hyperparameters and number of sampling points on PINN-1 and PINN-3 losses, a coarse grid search was performed, whereby, the number of neurons per hidden layer, number of hidden layers and number of sampling points in the domain $\mathcal{D}$ was varied. The different hyperparameters tested is shown in table 1, below. 

\begin{table} [h]
\caption{Selected hyperparameters} \vspace{2mm}
\centering
\begin{tabular}{ c c c } 
 \hline
 Parameters & PINN-1 & PINN-3 \\ \hline
 Num. neurons (all layers same) & 51, 86, 173 & 30, 50, 100 \\ 
 Num. layers & 5, 7, 9 & 5, 7, 9 \\ 
 Num. sampling points & 15e3, 30e3, 60e3 & 15e3, 30e3, 60e3 \\
 \hline
\end{tabular}
\end{table} 

The first hyperparameter sweep entails varying the number of neurons per hidden layer while keeping the number of layers and sampling points constant. To create a like-for-like comparison between the PINN-1 and PINN-3 models, the total number of training parameters per configuration was kept as close as possible. For example, the PINN-1 model with $51$ neurons per hidden layer (and 5 hidden layers) have $(2 \times 51) + 3 \cdot(51 \times 51) + (51 \times 18) = 8823$ trainable weight parameters and the PINN-3 model with $30$ neurons per hidden layer have $3 \cdot (2 \times 30) + 9 \cdot (30 \times 30) + (30 \times 6) + (30 \times 9) + (30 \times 9) = 8820$ weight parameters. To train the networks a decreasing learning rate schedule was used, similar to \cite{Cai2021}. The final results are obtained after 25000 epochs with learning rates of 5e-3 (5000 epochs), 1e-3 (10000 epochs) and 1e-4 (10000 epochs). Each network is trained three times for a given setting and the average results reported.
Figure 5 shows the average training history for the combined momentum, species and energy losses (PINN-1$\rightarrow$"Single" and PINN-3$\rightarrow$"Seg."). The results show that the combined loss values of both the PINN-1 and PINN-3 models reduce substantially during training, from approximately 1 to 1e-3. It is seen that the PINN-3 model produces lower losses compared to the PINN-1 approach for all cases, except the species and energy losses for the 51 neurons PINN-1 model which results in similar losses to that of the 30 neurons PINN-3 model. Averaging over all the training cases, the PINN-3 model has an 82\% lower momentum-, 87\% lower energy- and a 94\% lower species-loss compared to the PINN-1 losses. The best performing PINN-3 setting is 51 neurons and for the PINN-1 model, it is 86 neurons. The increased performance of the PINN-3 model over the PINN-1 model can be attributed to the reduction in the complexity of the transfer functions the three physics networks have to learn compared to the PINN-1 case where a single network is required to learn all the various physics.

Using the best-performing number of neurons per hidden layer settings mentioned above, the effect of MLP network depth on the combined loss values were investigated by retraining the models with an increasing number of hidden layers. Typically deeper networks require more training epochs to converge, therefore, the training schedule was adjusted as follows: 5e-3 (5000 epochs), 1e-3 (20000 epochs), 1e-4 (20000 epochs), 1e-5 (20000 epochs) and 1e-6 (20000 epochs). Figure 6 below shows the combined loss training history for different depth MLP networks. The results, again show the PINN-3 model producing lower losses for the momentum, species and energy predictions compared to the PINN-1 model results. The average percentage difference in momentum, energy and species loss values between the PINN models are 68\%, 78\% and 86\% respectively. The 7-layer PINN-3 model produced the lowest species and energy losses and the 5-layer variant produced the lowest momentum loss. The difference in losses between the 5- and 7-layer PINN-3 configurations is not large enough to merit the selection of the deeper network. The lowest PINN-1 losses were achieved using the 5-layer variant. Using these hyperparameter settings the effect of the number of sampling points on losses was also investigated. 
\begin{figure}[h!]
 \hspace*{-1.75cm}
 \includegraphics[scale=0.325]{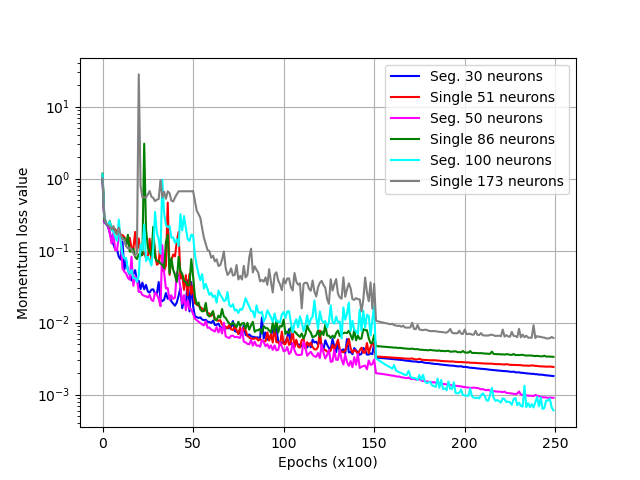}\includegraphics[scale=0.325]{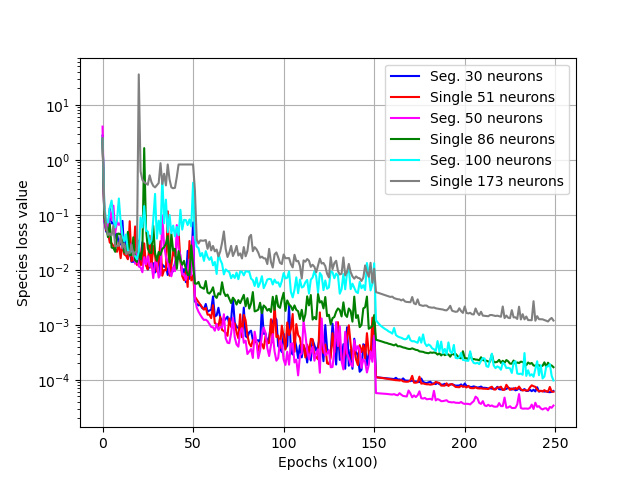}\includegraphics[scale=0.325]{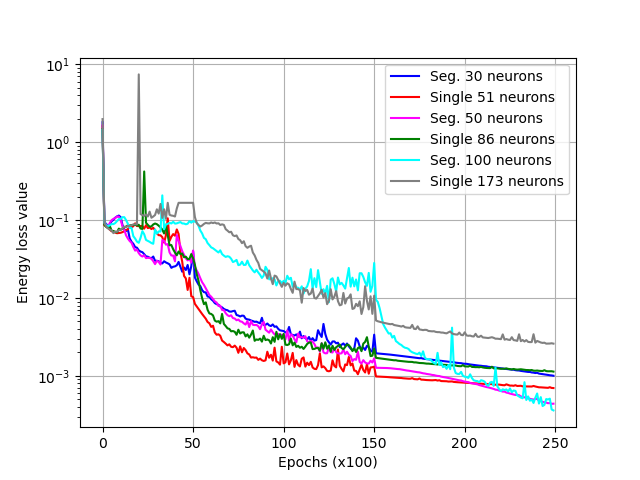}
\caption{Combined losses for varying number of hidden layer neurons}
\end{figure}
\begin{figure}[h!]
\centering
 \hspace*{-1.75cm}
 \includegraphics[scale=0.325]{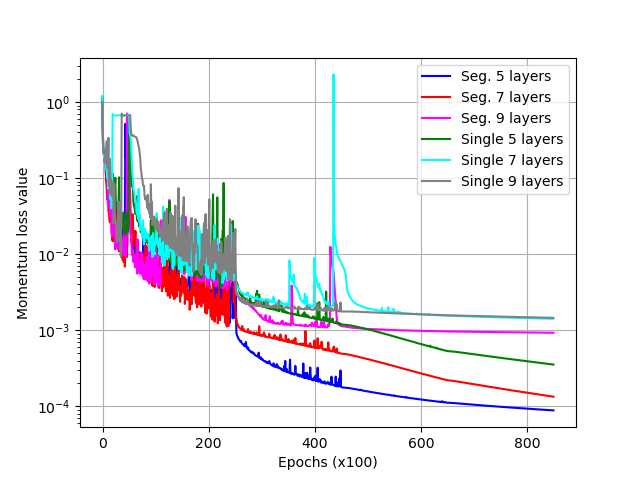}\includegraphics[scale=0.325]{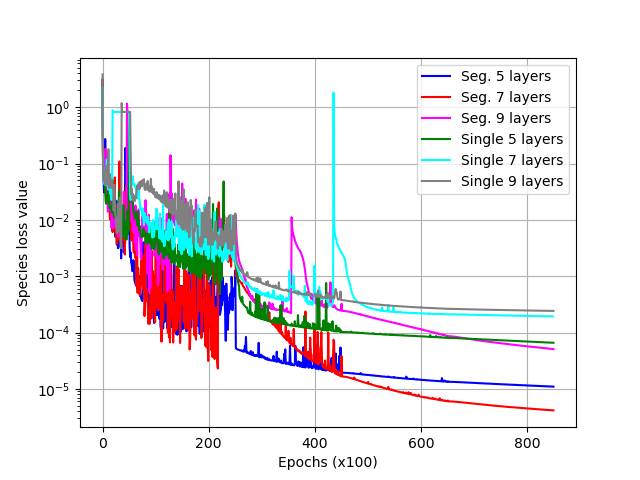}\includegraphics[scale=0.325]{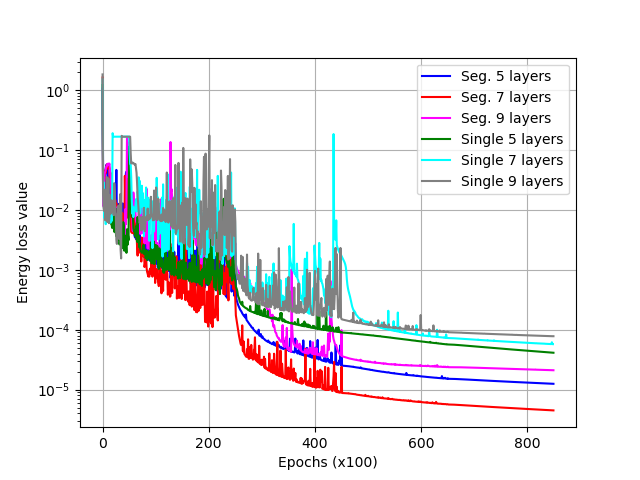}
\caption{Combined losses for varying number of hidden layers}
\end{figure}
\begin{figure}[h!]
\centering
 \hspace*{-1.75cm}
 \includegraphics[scale=0.325]{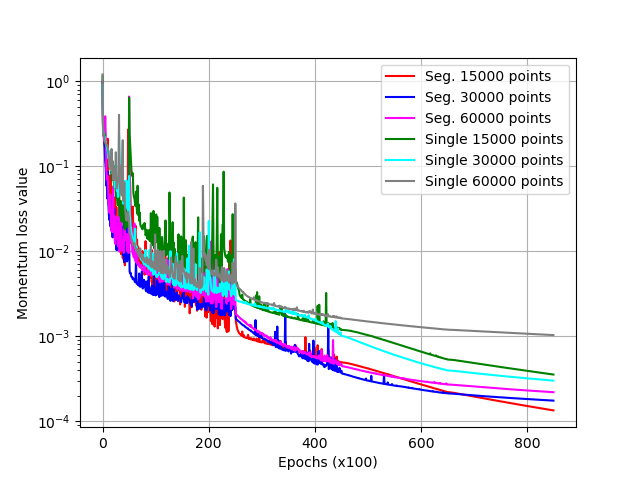}\includegraphics[scale=0.325]{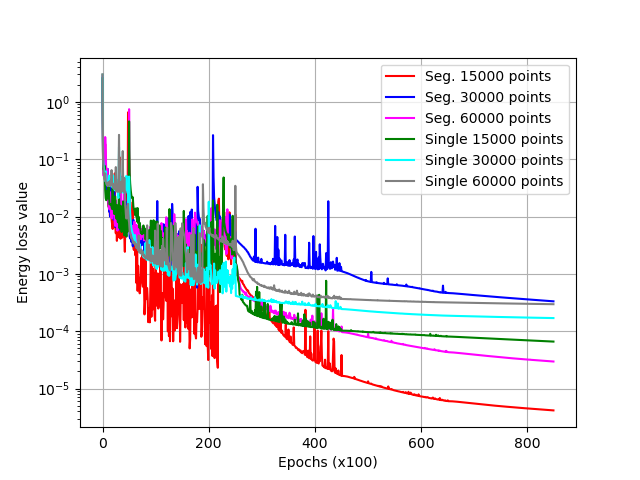}\includegraphics[scale=0.325]{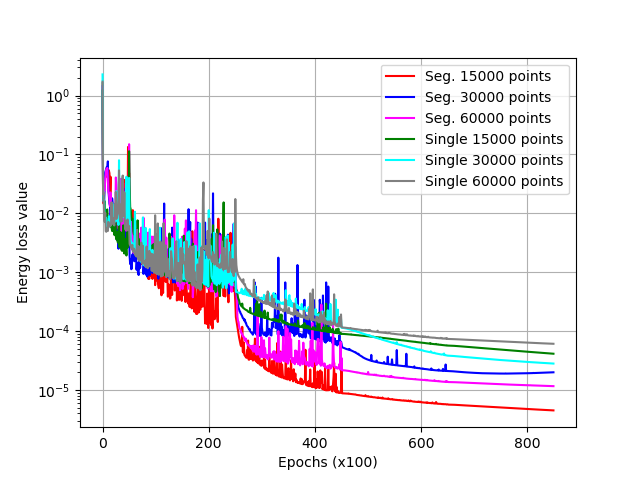}
\caption{Combined losses as a function of sampling density}
\end{figure}

Figure 7 shows the combined loss values for the PINN-1 ($5 \times 86$) and PINN-3 ($5 \times 50$) models trained with higher sampling densities. The graphs show that the PINN-3 model trained using 15000 sampling points outperforms the higher sampling density models. For the PINN-1 model, the best results were achieved training with 30000 coordinate points. It should be mentioned, that the larger models such as the 7- and 9-layer deep networks, theoretically, should produce lower losses compared to the smaller models, given that enough training epochs. But the current work, as will be shown later on, the training epochs and network size are sufficient to capture the solution of the physics PDEs. 

In the next section, the overall best-performing PINN-1 ($5 \times 86$, trained using 30000 points) and PINN-3 ($5 \times 50$ trained using 15000 points) temperature, velocity and species mass fraction predictions will be compared to the OpenFOAM simulation results.

\subsection{Comparison with CFD results}
Figure 8 shows the predicted X- and Y-velocity results of the CFD and PINN models. It is seen that the PINN-3 and CFD results are in good agreement, but the PINN-1 approach underpredicts the X- and Y-velocity profiles. To quantitatively compare the predictions of the PINNs with the CFD results, the normalised mean absolute percentage error (NMAPE) is calculated, as seen in \eqref{eq:22}.

\begin{equation} \label{eq:22}
	\text{NMAPE} = 100\% \times\frac{1}{N_i} \sum_{i=1}^{N_i} \frac{|\phi_{i,CFD} - \phi_{i,PINN}|}{\max(\phi_{i,CFD}) - \min(\phi_{i,CFD})}
\end{equation}

The NMAPEs for the X- and Y-velocity predictions by the PINN-1 model are 14.8\% and 5.6\% respectively. The velocity NMAPEs for the PINN-3 predictions are 8.2\% (X-velocity) and 2.5\% (Y-velocity). Figure 9 shows the predictions of the temperature field and species mass fractions. The temperature NMAPEs for the PINN-1 and PINN-3 models are 17\% and 8\% respectively. The NMAPEs for the PINN-1 and PINN-3 $H_2O$ mass fraction predictions are 16.5\% and 1.87\% and for the $O_2$ predictions 16.7\% and 1.86\% respectively.

\begin{figure}[h!]
\centering
 \includegraphics[scale=0.35]{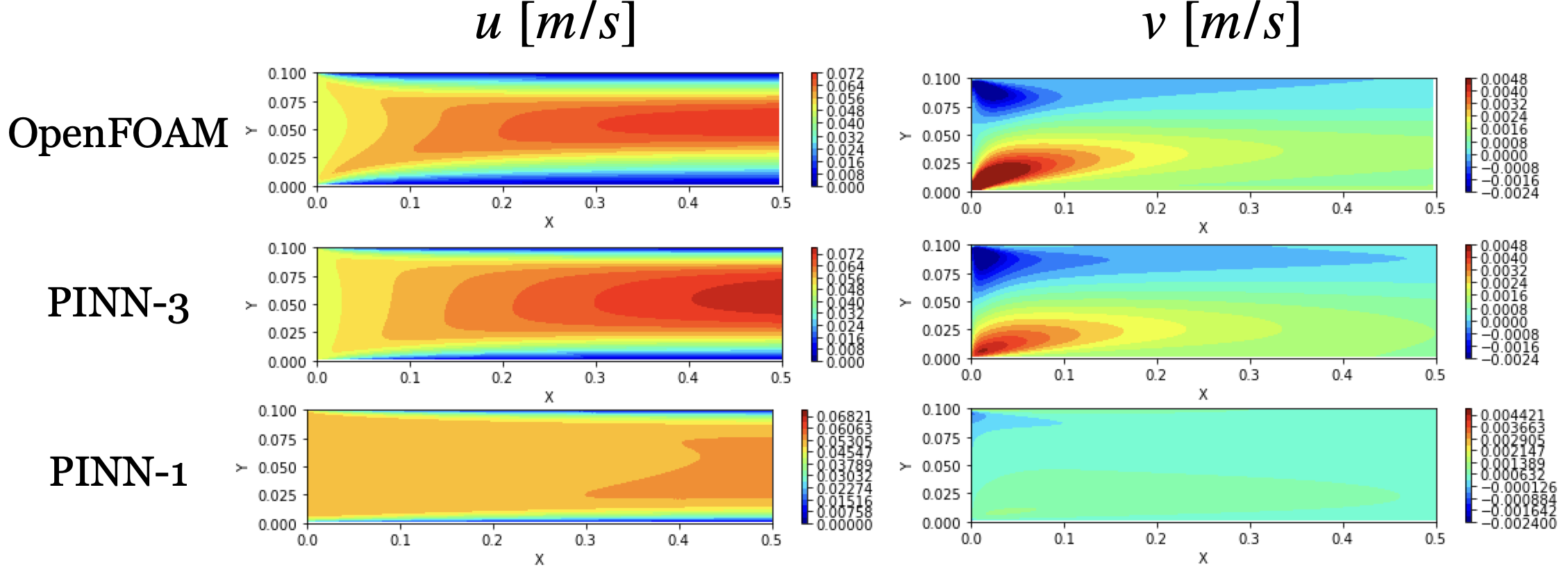}
\caption{Velocity predictions using PINN-1 ($5 \times 86$) and PINN-3 ($5 \times 50$)}
\end{figure}

\begin{figure}[h!]
\centering
 \includegraphics[scale=0.35]{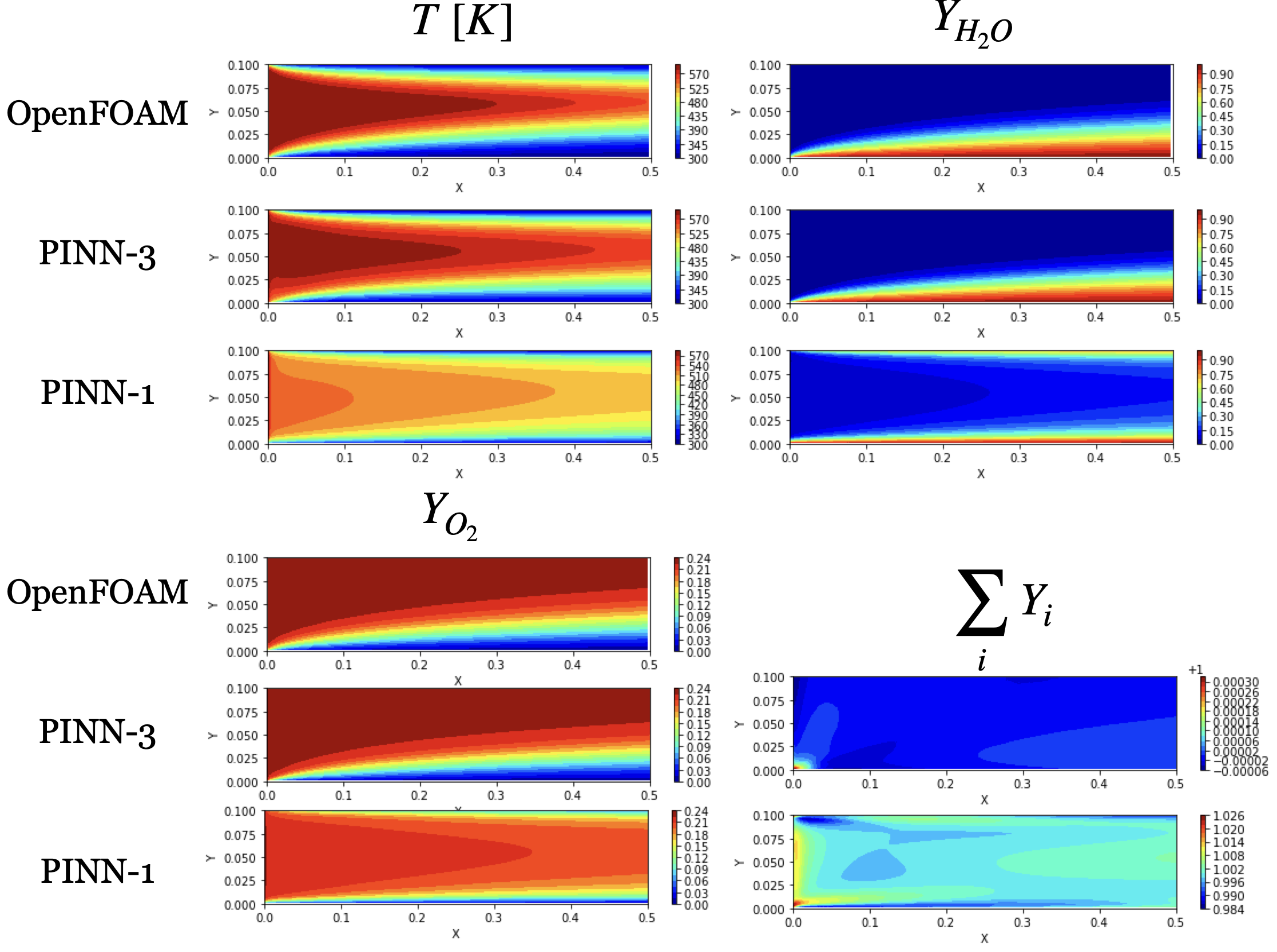}
\caption{Temperature and species mass fraction predictions using PINN-1 ($5 \times 86$) and PINN-3 ($5 \times 50$)}
\end{figure}

Figure 9 additionally shows the predictions of the summed mass fractions for the internal domain. As seen, the PINN-3 model results are all close to the desired value of 1, whereas, the PINN-1 approach struggles to enforce the $\mathcal{J}^{\sum Y_i}_r$ loss condition. A possible reason for the inability of the PINN-1 model to capture the desired physics could be that only a single combined loss value is minimised during training since only a single network is used. This leads to the minimisation of only certain aspects of the PDE equations. A solution to this is the application of active loss weighting during training, as shown by \cite{Niaki2020a}.

The $5 \times 86$ PINN-1 configuration was retrained multiple times to eliminate the chance that the optimiser was stuck in a local minimum due to the random starting location of the weights and biases.  But the retrained model predictions showed similar shortcomings. As an additional effort, the PINN-1 network width was reduced to 51 neurons and the model retrained. This yielded a significant increase in the accuracy of the velocity and temperature predictions. The X- and Y-velocity NMAPEs reduced to 7.2\% and 1.75\% respectively. The temperature NMAPEs value reduced to 6.5\%. These error percentage values made the PINN-1 model more accurate at predicting the temperature and velocity profiles when compared to the results of the $5 \times 50$ PINN-3 approach. The NMAPEs of the $H_2O$ and $O_2$ predictions by the $5 \times 51$ PINN-1 model are 6.0\% and 11.4\%, respectively, which is still higher than the PINN-3 predictions. The smaller PINN-1 model variant still struggled to ensure that the summed mass fractions at the internal sampling points add up to a value of 1. A possible reason why the narrower PINN-1 model outperforms the 86 neuron-wide PINN-1 variant could be because the former requires more training epochs to effectively learn the PDEs with a single MLP network, but this is still to be confirmed. 

The PINN-3 model was also retrained using 30000 sampling points to investigate if an increase in the sampling density improves the performance of the segregated PINN approach.  The increase in sampling density reduced the $5 \times 50$ PINN-3 model X- and Y-velocities NMAPEs to 7.5\% and 2.2\%. The temperature NMAPEs reduced to 7.8\% and the $H_2O$ and $O_2$ species mass fraction errors reduced to 1.5\% and 1.52\% respectively.

Further investigation into the predictions of the $5 \times 50$ PINN-3 and the $5 \times 51$ PINN-1 models trained using 30000 sampling points were conducted by plotting the velocity, temperature and species mass fraction profiles at different locations along the X-axis. Figure 10, below shows these profiles at X-axis locations of 0.1m, 0.25m and 0.4m. The velocity and temperature results show that both the PINN-1 (Single PINN) and PINN-3 (Seg. PINN) prediction trends are in good agreement with the CFD simulation (OF) results. The PINN-3 models slightly over predicts the temperatures in the lower half of the duct. The species mass fraction results show that the PINN-3 model can almost perfectly recreate the CFD simulation results, whereas, the PINN-1 results significantly deviate from the predicted CFD simulation profiles near the top wall. 

The results in figures 8-10 along with the mentioned error percentages shows that the PINN-3 architecture being a more robust and accurate modelling approach to capture the multi-species flow and heat transfer for the current problem under consideration. Seeing as the PINN-1 model can accurately solve for the momentum and energy PDEs, a possible solution to simplify the programming of the models and increase the momentum and energy solution accuracy is to have a single network resolving the momentum and energy PDEs and a single network resolving the species distributions.

\begin{figure}[h!]
\centering
 \includegraphics[scale=0.35]{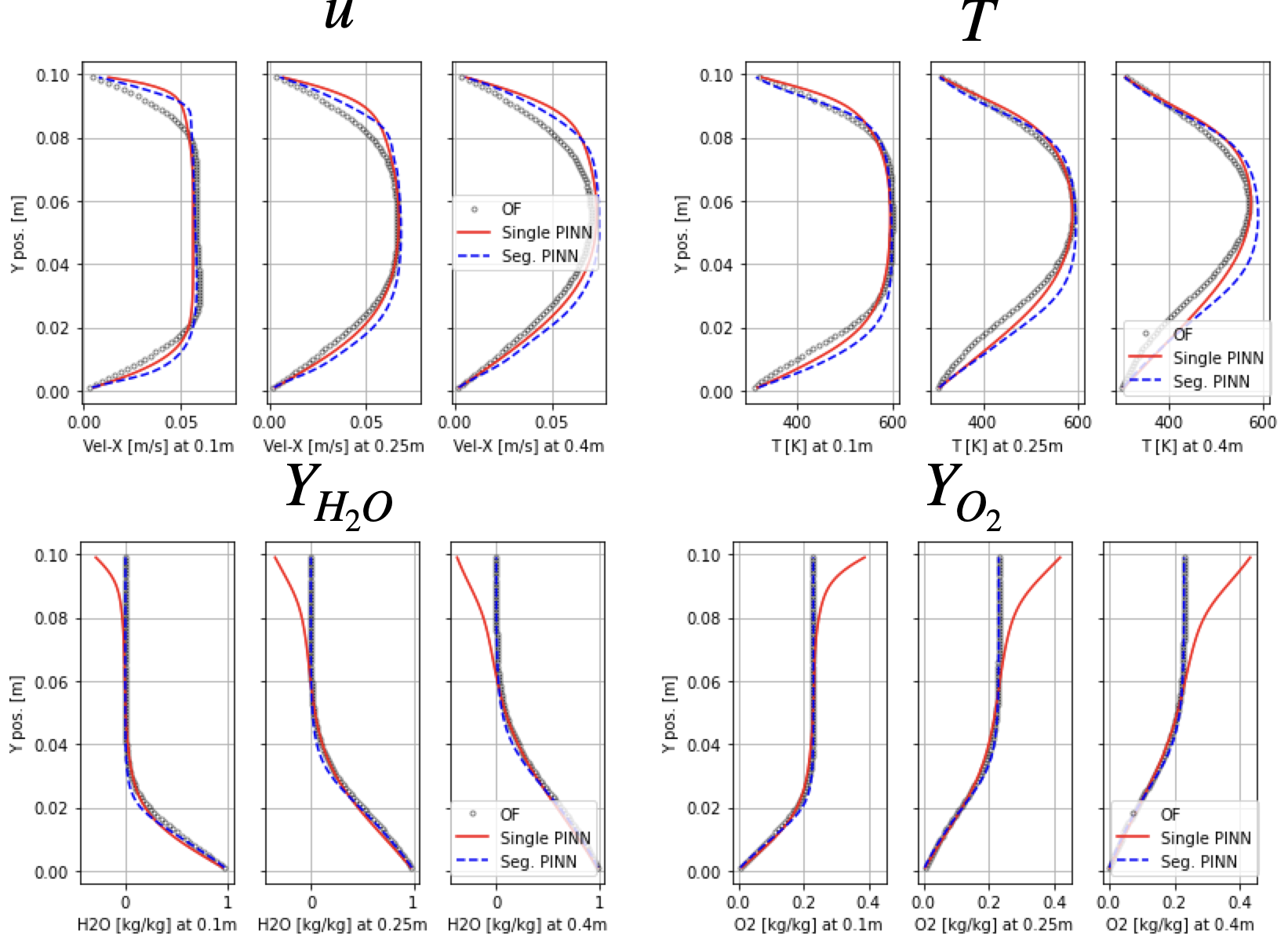}
\caption{Velocity, temperature and species mass fraction profiles for PINN-1 and PINN-3 models trained using 30000 sampling points}
\end{figure}

\subsection{Parameterised model}
To showcase the application of the PINN-3 architecture for surrogate modelling, a parameterised PINN of the 2D duct flow humidification problem is generated. This entails adding an additional MLP network input namely the water vapour boundary condition velocity $v_{bot}^{bc}$ and retraining the model using various sampling coordinates and boundary condition velocities. The range of velocity boundary conditions trained for was set to $0.0001 \rightarrow 0.01 \; [m/s]$. Figure 11 below shows the PINN-3 and OpenFOAM predictions for X-velocity, temperature and $H_2O$ mass fractions. For the three sets of results shown below the average velocity, temperature and water mass fraction NMAPEs are 7.7\%, 7.8\% and 2.43\% respectively.

\begin{figure}[h!]
\centering
 \hspace*{-1.75cm}
 \includegraphics[scale=0.3]{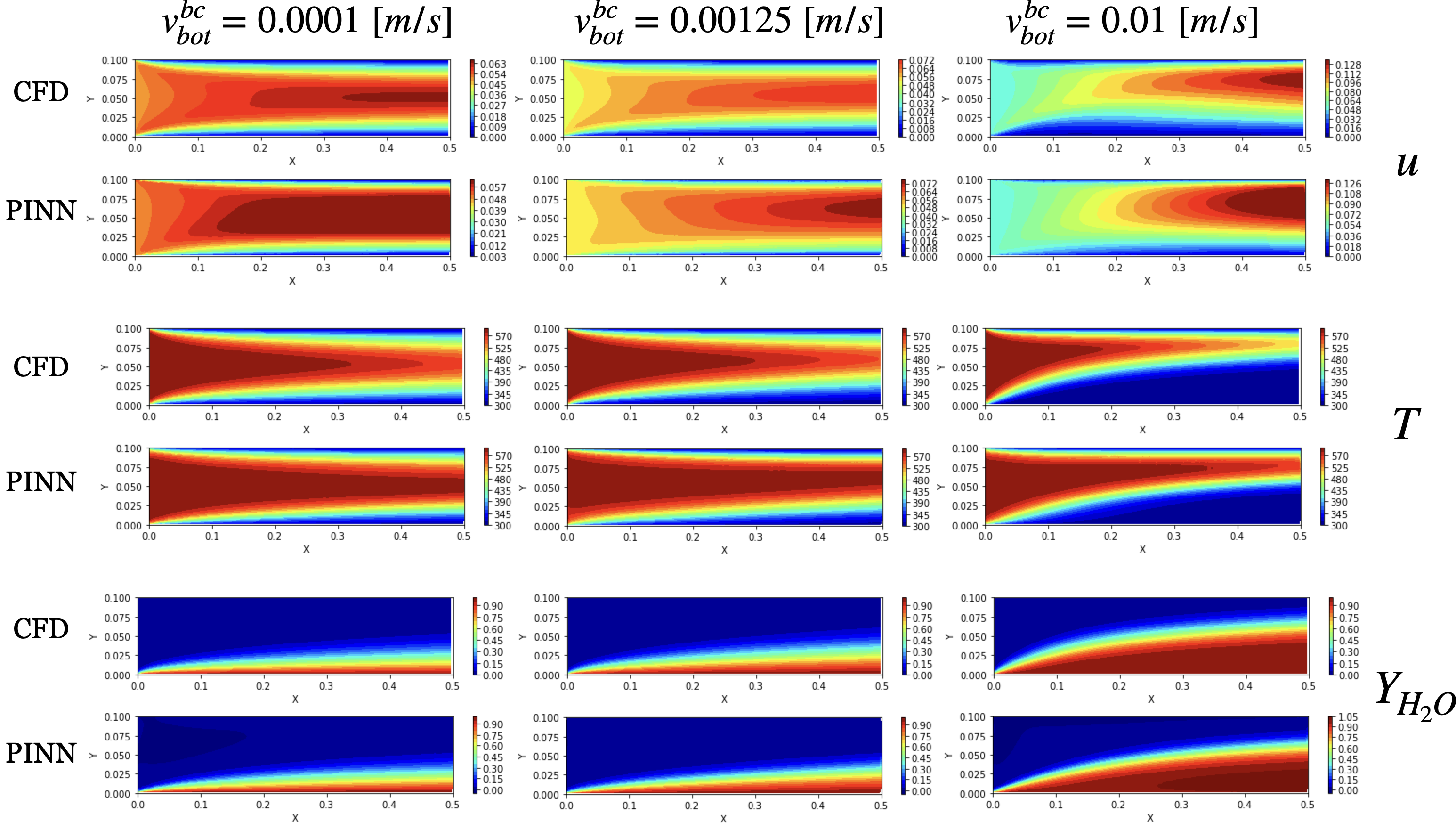}
\caption{Velocity, temperature and water mass fraction results for bottom wall inlet velocities of 0.0001, 0.00125 and 0.01 [m/s]}
\end{figure}

\section{Conclusions}
In the present work, two PINN modelling methodologies were implemented and used to solve simple multi-species flow and heat transfer on a 2D rectangular domain. The first approach uses a traditional PINN architecture to resolve all the desired physics PDEs, whereas, the second approach uses three separate neural networks each solving specific PDEs and their boundary conditions.

Models developed using the two PINN approaches were trained using different neural network architectures and sampling densities, to investigate the effect of hyperparameters on achievable combined loss values. The results showed that for nearly for all cases the PINN-3 approach yielded lower loss values compared to the resultant PINN-1 losses. On average for all the models trained, the PINN-3 losses were 62\% lower compared the PINN-1 values. The best performing PINN-1 and PINN-3 model predictions were also compared to the CFD simulation data for a single case. The results showed that the single network approach could adequately resolve the momentum and energy PDEs but struggled to enforce the species mass transport requirements, whereas, the PINN-3 approach successfully resolved the physics PDEs. A possible approach to reduce the amount of model setup time could be to use two networks, one for the momentum and energy and another for the species mass fraction predictions.

The PINN-3 approach was also applied to develop a surrogate model where the water vapour inlet velocity boundary conditions was an additional network input. The segregated PINN approach could adequately predict the velocity, temperature and water mass fraction distributions for various inlet velocities showcasing the ability to build parameterised models.

The PINN-3 approach applied in the current work is established as an attractive modelling methodology for PINNs when the network has to predict multiple (10+) solution variables, which can be grouped by different physics categories (species and energy transport).

Future work will look at expanding the segregated network approach to reactive flow problems, specifically combustion, and the addition of dynamic weighting of loss functions.

\bibliographystyle{ieeetr}
\bibliography{main.bib}
\end{document}